\documentclass[aps,prl,twocolumn,showpacs,groupedaddress]{revtex4}  
\usepackage{graphicx}  
\usepackage{dcolumn}   
\usepackage{bm}        
\usepackage{amssymb}   

\def\lsim{\mathrel{\rlap{\lower4pt\hbox{$\sim$}}
    \raise1pt\hbox{$<$}}}                

\newcommand{\GeV}       {\mbox{GeV}}

\newcommand{\invpb}     {\mbox{pb$^{-1}$}}
\newcommand{\pb}        {\mbox{pb}}

\newcommand{\MSB}     {\mbox{$m_{\tilde{b}}$}}

\newcommand{\met}    {\mbox{${\hbox{$E$\kern-0.6em\lower-.1ex\hbox{/}}}_T$}} 
\newcommand{\mht}    {\mbox{${\hbox{$H$\kern-0.75em\lower-.05ex\hbox{/}}}_T$}} 
\newcommand{\pt}    {\mbox{$p_T$}}

\newcommand{\Wemunujj}  {\mbox{$W(e\nu + \mu\nu)+j\bar{j}$}}

\newcommand{\Wtaunuj}   {\mbox{$W(\tau\nu)+ \ge 1~{\rm jet}$}}

\newcommand{\Znunujj}   {\mbox{$Z(\nu\bar{\nu})+j\bar{j}$}}

\newcommand{\Wemunubb}  {\mbox{$W(e\nu + \mu\nu)+b\bar{b}$}}

\newcommand{\Wtaunubb}  {\mbox{$W(\tau\nu)+b\bar{b}$}}
\newcommand{\Znunubb}   {\mbox{$Z(\nu\bar{\nu})+b\bar{b}$}}
\newcommand{\Znunucc}   {\mbox{$Z(\nu\bar{\nu})+c\bar{c}$}}

\newcommand{\dibosons}  {\mbox{$WW,WZ,ZZ$}}

\newcommand{\Wcj}  {\mbox{$W(e\nu+\mu\nu)+cj$}}

\begin{document}

\mbox{PRL, {\bf 97}, 171806 (2006)} \hspace{3.5in} \mbox{FERMILAB-PUB-06/269-E}

\title{Search for pair production of scalar bottom quarks
in $p\bar{p}$ collisions at $\bm{\sqrt{s}=}$1.96~TeV}

%
\author{                                                                      
V.M.~Abazov,$^{36}$                                                           
B.~Abbott,$^{76}$                                                             
M.~Abolins,$^{66}$                                                            
B.S.~Acharya,$^{29}$                                                          
M.~Adams,$^{52}$                                                              
T.~Adams,$^{50}$                                                              
M.~Agelou,$^{18}$                                                             
S.H.~Ahn,$^{31}$                                                              
M.~Ahsan,$^{60}$                                                              
G.D.~Alexeev,$^{36}$
% ...
\\                                                                            
\vskip 0.30cm                                                                 
\centerline{(D\O\ Collaboration)}                                             
\vskip 0.30cm                                                                 
}                                                                             
\affiliation{                                                                 
\centerline{$^{1}$Universidad de Buenos Aires, Buenos Aires, Argentina}       
\centerline{$^{2}$LAFEX, Centro Brasileiro de Pesquisas F{\'\i}sicas,         
                  Rio de Janeiro, Brazil}                                     
\centerline{$^{3}$Universidade do Estado do Rio de Janeiro,                   
                  Rio de Janeiro, Brazil}                                     
\centerline{$^{4}$Instituto de F\'{\i}sica Te\'orica, Universidade            
                  Estadual Paulista, S\~ao Paulo, Brazil}                     
\centerline{$^{5}$University of Alberta, Edmonton, Alberta, Canada,           
                  Simon Fraser University, Burnaby, British Columbia, Canada,}
\centerline{York University, Toronto, Ontario, Canada, and                    
                  McGill University, Montreal, Quebec, Canada}                
\centerline{$^{6}$Institute of High Energy Physics, Beijing,                  
                  People's Republic of China}                                 
\centerline{$^{7}$University of Science and Technology of China, Hefei,       
                  People's Republic of China}                                 
\centerline{$^{8}$Universidad de los Andes, Bogot\'{a}, Colombia}             
\centerline{$^{9}$Center for Particle Physics, Charles University,            
                  Prague, Czech Republic}                                     
\centerline{$^{10}$Czech Technical University, Prague, Czech Republic}        
\centerline{$^{11}$Center for Particle Physics, Institute of Physics,         
                   Academy of Sciences of the Czech Republic,                 
                   Prague, Czech Republic}                                    
\centerline{$^{12}$Universidad San Francisco de Quito, Quito, Ecuador}        
\centerline{$^{13}$Laboratoire de Physique Corpusculaire, IN2P3-CNRS,         
                   Universit\'e Blaise Pascal, Clermont-Ferrand, France}      
\centerline{$^{14}$Laboratoire de Physique Subatomique et de Cosmologie,      
                   IN2P3-CNRS, Universite de Grenoble 1, Grenoble, France}    
\centerline{$^{15}$CPPM, IN2P3-CNRS, Universit\'e de la M\'editerran\'ee,     
                   Marseille, France}                                         
\centerline{$^{16}$IN2P3-CNRS, Laboratoire de l'Acc\'el\'erateur              
                   Lin\'eaire, Orsay, France}                                 
\centerline{$^{17}$LPNHE, IN2P3-CNRS, Universit\'es Paris VI and VII,         
                   Paris, France}                                             
\centerline{$^{18}$DAPNIA/Service de Physique des Particules, CEA, Saclay,    
                   France}                                                    
\centerline{$^{19}$IPHC, IN2P3-CNRS, Universit\'e Louis Pasteur, Strasbourg,  
                    France, and Universit\'e de Haute Alsace,                 
                    Mulhouse, France}                                         
\centerline{$^{20}$Institut de Physique Nucl\'eaire de Lyon, IN2P3-CNRS,      
                   Universit\'e Claude Bernard, Villeurbanne, France}         
\centerline{$^{21}$III. Physikalisches Institut A, RWTH Aachen,               
                   Aachen, Germany}                                           
\centerline{$^{22}$Physikalisches Institut, Universit{\"a}t Bonn,             
                   Bonn, Germany}                                             
\centerline{$^{23}$Physikalisches Institut, Universit{\"a}t Freiburg,         
                   Freiburg, Germany}                                         
\centerline{$^{24}$Institut f{\"u}r Physik, Universit{\"a}t Mainz,            
                   Mainz, Germany}                                            
\centerline{$^{25}$Ludwig-Maximilians-Universit{\"a}t M{\"u}nchen,            
                   M{\"u}nchen, Germany}                                      
\centerline{$^{26}$Fachbereich Physik, University of Wuppertal,               
                   Wuppertal, Germany}                                        
\centerline{$^{27}$Panjab University, Chandigarh, India}                      
\centerline{$^{28}$Delhi University, Delhi, India}                            
\centerline{$^{29}$Tata Institute of Fundamental Research, Mumbai, India}     
\centerline{$^{30}$University College Dublin, Dublin, Ireland}                
\centerline{$^{31}$Korea Detector Laboratory, Korea University,               
                   Seoul, Korea}                                              
\centerline{$^{32}$SungKyunKwan University, Suwon, Korea}                     
\centerline{$^{33}$CINVESTAV, Mexico City, Mexico}                            
\centerline{$^{34}$FOM-Institute NIKHEF and University of                     
                   Amsterdam/NIKHEF, Amsterdam, The Netherlands}              
\centerline{$^{35}$Radboud University Nijmegen/NIKHEF, Nijmegen, The          
                  Netherlands}                                                
\centerline{$^{36}$Joint Institute for Nuclear Research, Dubna, Russia}       
\centerline{$^{37}$Institute for Theoretical and Experimental Physics,        
                   Moscow, Russia}                                            
\centerline{$^{38}$Moscow State University, Moscow, Russia}                   
\centerline{$^{39}$Institute for High Energy Physics, Protvino, Russia}       
\centerline{$^{40}$Petersburg Nuclear Physics Institute,                      
                   St. Petersburg, Russia}                                    
\centerline{$^{41}$Lund University, Lund, Sweden, Royal Institute of          
                   Technology and Stockholm University, Stockholm,            
                   Sweden, and}                                               
\centerline{Uppsala University, Uppsala, Sweden}                              
\centerline{$^{42}$Physik Institut der Universit{\"a}t Z{\"u}rich,            
                   Z{\"u}rich, Switzerland}                                   
\centerline{$^{43}$Lancaster University, Lancaster, United Kingdom}           
\centerline{$^{44}$Imperial College, London, United Kingdom}                  
\centerline{$^{45}$University of Manchester, Manchester, United Kingdom}      
\centerline{$^{46}$University of Arizona, Tucson, Arizona 85721, USA}         
\centerline{$^{47}$Lawrence Berkeley National Laboratory and University of    
                   California, Berkeley, California 94720, USA}               
\centerline{$^{48}$California State University, Fresno, California 93740, USA}
\centerline{$^{49}$University of California, Riverside, California 92521, USA}
\centerline{$^{50}$Florida State University, Tallahassee, Florida 32306, USA} 
\centerline{$^{51}$Fermi National Accelerator Laboratory,                     
            Batavia, Illinois 60510, USA}                                     
\centerline{$^{52}$University of Illinois at Chicago,                         
            Chicago, Illinois 60607, USA}                                     
\centerline{$^{53}$Northern Illinois University, DeKalb, Illinois 60115, USA} 
\centerline{$^{54}$Northwestern University, Evanston, Illinois 60208, USA}    
\centerline{$^{55}$Indiana University, Bloomington, Indiana 47405, USA}       
\centerline{$^{56}$University of Notre Dame, Notre Dame, Indiana 46556, USA}  
\centerline{$^{57}$Purdue University Calumet, Hammond, Indiana 46323, USA}    
\centerline{$^{58}$Iowa State University, Ames, Iowa 50011, USA}              
\centerline{$^{59}$University of Kansas, Lawrence, Kansas 66045, USA}         
\centerline{$^{60}$Kansas State University, Manhattan, Kansas 66506, USA}     
\centerline{$^{61}$Louisiana Tech University, Ruston, Louisiana 71272, USA}   
\centerline{$^{62}$University of Maryland, College Park, Maryland 20742, USA} 
\centerline{$^{63}$Boston University, Boston, Massachusetts 02215, USA}       
\centerline{$^{64}$Northeastern University, Boston, Massachusetts 02115, USA} 
\centerline{$^{65}$University of Michigan, Ann Arbor, Michigan 48109, USA}    
\centerline{$^{66}$Michigan State University,                                 
            East Lansing, Michigan 48824, USA}                                
\centerline{$^{67}$University of Mississippi,                                 
            University, Mississippi 38677, USA}                               
\centerline{$^{68}$University of Nebraska, Lincoln, Nebraska 68588, USA}      
\centerline{$^{69}$Princeton University, Princeton, New Jersey 08544, USA}    
\centerline{$^{70}$State University of New York, Buffalo, New York 14260, USA}
\centerline{$^{71}$Columbia University, New York, New York 10027, USA}        
\centerline{$^{72}$University of Rochester, Rochester, New York 14627, USA}   
\centerline{$^{73}$State University of New York,                              
            Stony Brook, New York 11794, USA}                                 
\centerline{$^{74}$Brookhaven National Laboratory, Upton, New York 11973, USA}
\centerline{$^{75}$Langston University, Langston, Oklahoma 73050, USA}        
\centerline{$^{76}$University of Oklahoma, Norman, Oklahoma 73019, USA}       
\centerline{$^{77}$Oklahoma State University, Stillwater, Oklahoma 74078, USA}
\centerline{$^{78}$Brown University, Providence, Rhode Island 02912, USA}     
\centerline{$^{79}$University of Texas, Arlington, Texas 76019, USA}          
\centerline{$^{80}$Southern Methodist University, Dallas, Texas 75275, USA}   
\centerline{$^{81}$Rice University, Houston, Texas 77005, USA}                
\centerline{$^{82}$University of Virginia, Charlottesville,                   
            Virginia 22901, USA}                                              
\centerline{$^{83}$University of Washington, Seattle, Washington 98195, USA}  
}                                                                             
\date{August 7, 2006}

\begin{abstract}
A search for direct production of scalar bottom quarks~($\tilde{b}$)
is performed with 310~\mbox{pb$^{-1}$} of data collected by the D\O\ experiment
in~$p\bar{p}$ collisions at~\mbox{$\mathrm{\sqrt{s}=1.96~TeV}$} 
at the Fermilab Tevatron Collider. The topology analyzed consists of 
two $b$ jets and an imbalance in transverse momentum due to undetected
neutralinos~($\tilde{\chi}^0_1$), with $\tilde{\chi}^0_1$ assumed to be the 
lightest supersymmetric particle. 
We find the data consistent with standard model 
expectations, and set a 95\%~C.L. exclusion domain in the 
($m_{\tilde{b}},m_{\tilde{\chi}^0_1}$) mass plane, improving significantly 
upon the results from Run I of the Tevatron. 

\end{abstract}

\maketitle 

Supersymmetric ({\tt SUSY}) models~\cite{theo:SUSY} provide an extension of 
the standard model ({\tt SM}) with mechanisms viable for the unification 
of interactions and a solution to the hierarchy problem.
Particularly attractive are models that conserve {\it R}-parity, in which
{\tt SUSY} particles are produced in pairs and the lightest supersymmetric 
particle ({\tt LSP}) is stable.
In {\tt SUSY}, a scalar field is associated to each of the left and 
right handed chirality states of a given {\tt SM} quark or lepton. Two 
mass eigenstates result from the mixing of these scalar fields.
The spin-1/2 partners of the neutral gauge and Higgs bosons are called
neutralinos.  

In supergravity inspired models~\cite{theo:sugra},
the lighest neutralino $\tilde{\chi}^0_1$ arises as the natural
{\tt LSP}, and, being neutral and weakly interacting, could be responsible
for the dark matter in the universe.
For large values of $\tan \beta$ (the ratio of the vacuum
expectation values of the two Higgs fields) the mixing term among the
scalar fields associated with the bottom quark is large. Therefore, a 
large splitting is expected among the mass eigenstates, 
that could result in a low mass value for one of them, 
hereafter called scalar bottom quark or sbottom ($\tilde{b}$).
The {\tt SUSY} particle mass hierarchy can even be such 
that the decay $\tilde{b}\to b\tilde{\chi}^0_1$ is the only one kinematically 
allowed~\cite{theo:sblight}, an assumption that is made in the following.
 
In this Letter, a search is reported for $\tilde{b}$ pair production with 
310~\mbox{pb$^{-1}$} of data collected 
during Run II of the Fermilab Tevatron.
At leading order, the $\tilde{b}$ pair production cross section in $p\bar p$ 
collisions depends only on the sbottom mass. For a center-of-mass
energy $\sqrt{s}=1.96$~TeV,
the next-to-leading order ({\tt NLO}) cross section, calculated with 
{\sc prospino-2}~\cite{ref:prospino2} ranges from 15 to 0.084 \pb\ 
for sbottom masses between 100 and 230~\GeV, with very little dependence 
on the masses of the other SUSY particles.
The final state of this process corresponds to  
two $b$~jets and missing transverse energy (\met) due to the undetected 
neutralinos. 
The maximum sbottom mass (\MSB) excluded by previous results is 148~\GeV~\cite{theo:D0-CDF-LEP-sb}. 

A full description of the D\O\ detector is available in 
Ref.~\cite{Abazov:2005pn}. The central tracking system consists of a
silicon microstrip tracker and a central fiber tracker,
both located within a 1.9~T superconducting solenoid. A liquid-argon 
and uranium calorimeter covers pseudorapidities up to $|\eta|$ $\approx 4.2$, where 
$\eta=-\ln \left[ \tan \left( \theta/2 \right) \right]$ and $\theta$ is the polar angle 
relative to the proton beam. The calorimeter has three sections, 
housed in separate cryostats: the central one covers $|\eta|$ $\lsim 1.1$, and the two end 
sections extend the coverage to larger $\vert\eta\vert$. The calorimeter is segmented in depth, 
with four electromagnetic layers followed by up to five hadronic layers. It is also segmented 
into projective towers of $0.1\times 0.1$ size in $\eta - \phi$ space, where $\phi$ is the azimuth 
in radians.  
An outer muon system, covering $|\eta|<2$, consists of a layer of tracking detectors and scintillation 
trigger counters positioned in front of 1.8~T toroids, followed by two similar layers
after the toroids. Jet reconstruction is based on the Run~II cone algorithm~\cite{jetalgo} with a 
cone size of 0.5, that uses energies deposited in calorimeter towers.
Jet energies are calibrated using transverse momentum balance in photon+jet 
events. The missing transverse energy in an event is based on all calorimeter cells, 
and is corrected for the jet energy calibration and for reconstructed muons.

The D\O\ trigger has three levels: L1, L2, and L3. 
The data were collected with triggers specifically 
designed for \met +jets topologies. We define $\mht = |\sum_{\rm jets} \vec{\pt}|$ 
the vector sum of the jet transverse 
momenta. The trigger conditions at L1 require at least three calorimeter towers
with $E_T>5$~\GeV, where a trigger tower spans 
$\Delta\phi\times\Delta\eta = 0.2\times 0.2$. We then require $\mht>$~20~(30)~\GeV\ at L2 (L3).
Approximately 14~million events were collected with the \met +jets triggers.

The signal is simulated in the framework of a generic minimal supersymmetric standard model,
in which we vary the masses of the $\tilde{b}$ and $\tilde{\chi}_1^0$, all other 
parameters being fixed. 
The masses of the other {\tt SUSY} particles are set such that the only sbottom 
decay mode is into $b\tilde{\chi}^0_1$.
The {\tt SUSY} and {\tt SM} processes are processed using Monte Carlo ({\tt MC}) generators
{\sc pythia 6.202}~\cite{ref:pythia} for the signal, {\sc alpgen 1.3.3}~\cite{ref:alpgen} interfaced 
with {\sc pythia} for the {\tt SM}. All the events are passed through
a full {\sc geant-3}~\cite{ref:geant3} simulation of the geometry and response of the D\O\ 
detector with an average of 0.8 minimum-bias events overlayed on each generated event.
The {\tt CTEQ5L} parton density functions ({\tt PDF})~\cite{ref:pdflib} are used in the simulation.

Instrumental background from mismeasurement of jet energies in
multijet events is estimated from data, and is referred to as ``{\tt QCD}'' background in the following.
The main {\tt SM} backgrounds relevant to our analysis are from vector boson
production in association with jets, and top quark production.
To estimate the backgrounds from $W/Z$+jets processes, we use the {\tt NLO} cross sections computed with 
{\sc mcfm}~\cite{ref:mcfm}. 
The theoretical {\tt NNLO} $t\bar t$ production cross section 
is taken from Ref.~\cite{ref:tt-cross-section}. 

The events are examined to ensure 
that the reconstructed vertex corresponds to the actual position of the primary vertex ({\tt PV}).
We select events that are well contained in the detector by restricting the {\tt PV} 
within 60~cm along the beam direction with respect to the detector center.
We define a charged-particle fraction ({\tt CPF}) as the ratio of the
charged-particle transverse energy,
computed from the sum of scalar \pt\ values of charged particles
(reconstructed in the tracking system) that emanate from the {\tt PV}
and are associated with a jet, divided by the jet transverse energy
measured in the calorimeter. The two leading jets, i.e those with the largest transverse energies, 
are required to have {\tt CPF}$>$0.05. 
This criterion rejects events with fake jets or where a wrong {\tt PV} is selected.
The overall inefficiency associated with this procedure is measured using events 
collected at random beam crossings, and events with two
jets emitted back-to-back in azimuth.
The jets must also have energy fraction in the electromagnetic layers of the calorimeter~$<$0.95
and \pt$>30,15$~\GeV\ for the first and second leading jets. This set of initial
cuts requires in addition $\Delta\phi <$~165$^\circ$, where
$\Delta\phi$ is defined as the difference in azimuth between the
two leading jets. 

Table~\ref{tab3:Cuts1} defines our selection criteria, and shows the effect of applying them sequentially
in the analysis of data, and their impact on signal efficiency, for the choice of
($m_{\tilde{b}}$,$m_{\tilde{\chi}_1^0}$) = (140,80)~\GeV. Criteria {\bf C1}-{\bf C4} are effective 
against {\tt QCD}, {\bf C2} and {\bf C4}-{\bf C8} against vector-bosons+jets, and {\bf C9} suppresses 
$t\bar t$ background. 
For $\tilde{b}$ masses of $\sim 100$~\GeV, the mean \met\ and jet \pt\ are close to 
what is expected from {\tt SM} backgrounds, but are substantially larger for higher $\tilde{b}$ masses.
The selections are tuned on {\tt MC} so as to maintain good sensitivity to signal 
for small $\tilde{b}$ masses, using minimal values for threshold requirements, for instance 
\met$>$60~\GeV\ ({\bf C1}) and \pt$>40,20$~\GeV\ ({\bf C2}) for the first and second leading jets. 
Later we show that, depending on the masses ($m_{\tilde{b}}$,$m_{\tilde{\chi}_1^0}$), 
higher thresholds on \met\ and jet~\pt\ can be applied to 
increase the sensitivity to signal.

The first and second leading jets are required to be in the central
region of the calorimeter, $|\eta^{\rm det}|<$1.1~and~2.0 respectively ({\bf C3}), 
where $\eta^{\rm det}$ is the jet pseudorapidity calculated with a jet origin at the detector center.
Because of the central production of $\tilde{b}$ events, these selections do not affect signal
efficiency, but reduce background.
We also define $\Delta\phi_{min}$(\met,jets) and $\Delta\phi_{max}$(\met,jets), the minimum and maximum of the 
differences in azimuth between the direction of \met\ and the direction of any jet. 
Requiring $\Delta\phi_{min}>$35$^\circ$ rejects {\tt QCD} events ({\bf C4}),
and $\Delta\phi_{min}<$120$^\circ$ and $\Delta\phi_{max}<$175$^\circ$ suppress 
{\tt SM} background ({\bf C4, C5}).

Since we do not expect isolated electrons, muons or tau leptons in signal events,
vetoes are imposed on events with an isolated electron ({\bf C6}), muon ({\bf C7}), or a charged track ({\bf C8}) with 
\pt$>$5~\GeV. 
Electrons and muons are defined isolated based on a criterion for energy deposition in a cone around the lepton
direction in the calorimeter. A charged track is considered 
isolated if no other charged track with \pt$>$1.5~\GeV\ is found in a hollow cone with inner and outer radii 0.05 and 0.2,
formed around the direction of the track. The last requirement ({\bf C9}) stipulates that either two or three 
jets are allowed.

\begin{table}[hbtp]
\caption{\label{tab3:Cuts1}
Sequence of criteria applied for the selection of events with their 
corresponding impact on data and on signal efficiency (Eff.) for 
($m_{\tilde{b}}$,$m_{\tilde{\chi}_1^0}$) = (140,80).}
\begin{ruledtabular}
\begin{tabular}{llrr}
\multicolumn{2}{l}{Selection criterion applied}                           & Events left  & Eff. (\%) \\
\hline
{\bf C1 :}      & $\met>60$~\GeV                                         & 16,279 & 18  \\
{\bf C2 :}      & $\pt_1>40$~\GeV, $\pt_2>20$~\GeV                       & 14,095 & 16  \\
{\bf C3 :}      & $|\eta^{\rm det}_{\rm jet1}|<1.1$, $|\eta^{\rm det}_{\rm jet2}|<2.0$ &  9,653 & 14  \\
{\bf C4 :}      & $ 35^\circ<\Delta \phi_{min}(\met,\rm{jets})<120^\circ$     &  3,149 & 10  \\
{\bf C5 :}      & $\Delta \phi_{max}(\met,\rm{jets}) < 175^\circ$             &  2,783 &  9   \\
{\bf C6 :}      & {\rm isolated electron veto}                                          &  2,059 &  9   \\
{\bf C7 :}      & {\rm isolated muon veto}                                              &  1,809 &  9   \\
{\bf C8 :}      & {\rm isolated track veto}                                            &    756 &  7   \\
{\bf C9 :}      & 2 {\rm or} 3 {\rm jets}                                            &    671 &  6   \\
\end{tabular} 
\end{ruledtabular}
\end{table}

Table~\ref{tab:NumberOfEventBeforeAfterBtagging} 
gives the numbers of events expected for {\tt SM}
backgrounds and signal, and the number of events observed in data after the above selections.
Since an important fraction of the background corresponds to processes 
with light-flavor jets in the final state, we take advantage of the presence
of $b$ jets in the signal to significantly increase the sensitivity of the search by 
using a lifetime-based heavy-flavor tagging algorithm ($b$-tagging).
Based on the impact parameters of the tracks in the jet, the algorithm~\cite{ref:jlip} 
computes a probability for a jet to be light-flavored.

We select the $b$-tagging probability such that 0.1\% of the light-flavored jets
are tagged for jets having \pt\ of 50~\GeV\ as yielding the best expected signal sensitivity. 
The corresponding typical tagging efficiencies for $c$- and $b$-quark jets are 5\% and 30\%, respectively. 
Because the current detector simulation does not reproduce the tracking precisely enough, the b-tagging 
algorithm is not applied to simulated jets directly. Instead, jets are weighted by their 
probability to be $b$-tagged, according to their flavor, using parameterizations derived from data.
In what follows, we require at least one $b$-tagged jet in the event.
Requiring more than one $b$-tagged jet would lower slightly the sensitivity of the analysis.
\begin{table}[hbtp]
\caption{\label{tab:NumberOfEventBeforeAfterBtagging}
Numbers of events expected from {\tt SM} and {\tt QCD} backgrounds, of data events observed, 
and of signal events expected, after all selection criteria, both before ($N_{\text exp}$) and after $b$-tagging.
All uncertainties are statistical only. Backgrounds from $b$, $c$, and light jets ($j$) are shown separately.}
\begin{ruledtabular}
\begin{tabular}{lrr}
{\tt SM} process      & \multicolumn{1}{r}{$N_{\text exp}$}         &  \multicolumn{1}{r}{with $b$-tagging} \\
\hline
\Wemunujj             & 155    $\pm$  13    & 1.9  $\pm$ 0.2             \\ 
\Wcj                  &   2.2  $\pm$  0.6   & 0.2  $\pm$ 0.1     \\ 
\Wemunubb             &   1.1  $\pm$  0.1   & 0.6  $\pm$ 0.1             \\ 
\Wtaunuj              & 101    $\pm$  14    & 4.1  $\pm$ 0.6     \\ 
\Wtaunubb             &   2.2  $\pm$  0.3   & 1.0  $\pm$ 0.1     \\ 
\Znunujj              & 257    $\pm$  12    & 3.9  $\pm$ 0.2     \\ 
\Znunucc              &   8.0  $\pm$  0.7   & 0.9  $\pm$ 0.1     \\ 
\Znunubb              &   7.8  $\pm$  0.3   & 4.0  $\pm$ 0.2     \\ 
\dibosons             &  14.2  $\pm$  0.7   & 0.9  $\pm$ 0.2             \\
{\rm top production}&  7.9  $\pm$  0.2   & 3.8  $\pm$ 0.2                  \\
\hline                                            
Total {\tt SM}        & 556    $\pm$  23    & 21.5 $\pm$ 0.8    \\ 
\hline                                            
{\tt QCD} background & 109  $\pm$ 9      &  4   $\pm$ 2        \\  
\hline                                            
Data                  & 671              & 22                  \\
\hline
$(m_{\tilde{b}},m_{\tilde{\chi}_1^0}) = (140,80)$~GeV & 43 $\pm$ 2 & 23.1 $\pm$ 0.9 \\ 
\end{tabular}
\end{ruledtabular}
\end{table}

\begin{figure}[hbtp]
\begin{minipage}{\linewidth}
  \begin{center}
   \includegraphics[width=1.\linewidth]
{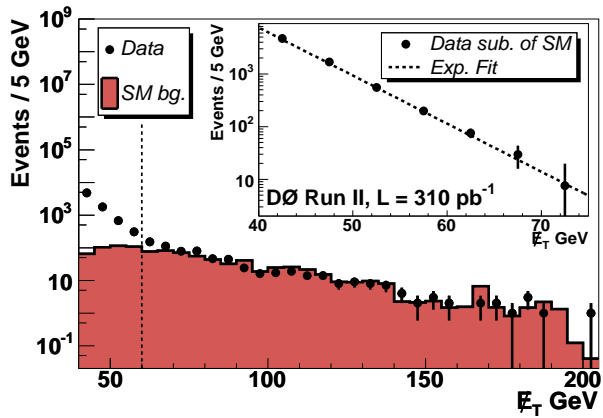}
  \end{center}
   \caption{Distribution in \met\ after applying all criteria, except \met~$> 60$~\GeV\ ({\bf C1}). 
The dark shaded area corresponds to the {\tt SM} simulation. A fit by an exponential to \met$<60$~\GeV, after 
subtraction of the contributions from the {\tt SM}, shown in the figure insert, is used to estimate the
instrumental background.}   
    \label{fig:ShowMETfit.eps}
\end{minipage}
\end{figure}

In order to estimate the background from {\tt QCD}, we compare 
our selected data sample, without imposing the criterion on \met~({\bf C1}), to the simulation 
of background from {\tt SM}.
Figure~\ref{fig:ShowMETfit.eps} shows that data are well reproduced by the {\tt SM} background
at high \met. We therefore attribute the exponential rise at low \met\ to {\tt QCD} multijet instrumental background.
A fit by an exponential to the data for \met~$< 60$~\GeV, after subtraction of the contributions from the {\tt SM}, is
shown in the insert in Fig.~\ref{fig:ShowMETfit.eps}. When the fit is
extrapolated to \met~$> 60$~\GeV, it provides an estimate of 109$\pm$9 {\tt QCD} events.
After $b$-tagging, this procedure estimates the presence of only 4$\pm$2 events. 
Given the larger \met\ threshold we use for higher sbottom masses, we expect that, 
after the $b$-tagging, less than two {\tt QCD} events will survive the final event selection.
The {\tt QCD} contribution is therefore neglected in the rest of this analysis.  
Table~\ref{tab:NumberOfEventBeforeAfterBtagging} 
shows the results after all selections, including $b$-tagging, for {\tt SM} backgrounds, data and signal.

As already mentioned, the mean \met\ and jet \pt\ become substantially larger for higher 
sbottom masses than the values expected from {\tt SM} backgrounds.
This provides a handle for improving the sensitivity to the signal for large $m_{\tilde{b}}$. 
Table~\ref{tab:high_mass} shows results for two higher sbottom-mass points, the chosen \met\ and \pt\ thresholds, 
together with the resulting number of events found after all selections, including $b$-tagging,
for data, {\tt SM} background and signal. For the highest sbottom masses probed, we note a deficit 
in the number of events observed compared to the SM background expectation. The probability of such a deficit is 4\%.

\begin{table}[hbtp]
\caption{\label{tab:high_mass} Optimized values for the criteria {\bf C1} and {\bf C2}, numbers 
of data events observed, numbers of events expected from {\tt SM} and signal  
for two ($m_{\tilde{b}}$,$m_{\tilde{\chi}_1^0}$) masses after $b$-tagging (statistical uncertainties only).}
\begin{ruledtabular}
\begin{tabular}{lcc}
($m_{\tilde{b}}$,$m_{\tilde{\chi}_1^0}$) in \GeV   
                                                                 & (180,90)         &    (215,0)      \\
\hline
{\bf C1}: \met~[\GeV]                                            &  60              &      80         \\
{\bf C2}: jet 1 \pt~[\GeV]                                       &  70              &     100         \\
{\bf C2}: jet 2 \pt~[\GeV]                                      &  40              &      50         \\
\hline
data                                                             &  7               &       0         \\
{\tt SM}                                                         & $8.9 \pm 0.3$    & $3.2 \pm 0.2$  \\
signal                                                           & $9.4 \pm 0.3$    & $4.6 \pm 0.1$  \\
\end{tabular}
\end{ruledtabular}
\end{table}

The following systematic uncertainties are taken into account in deriving the final results.
The integrated luminosity contributes an uncertainty of 6.5\%. 
The uncertainty from jet energy calibration is typically of the order of 7\%.
The total uncertainty from jet energy resolution, 
jet track confirmation, misvertexing and jet reconstruction is 5\%. 
The systematic uncertainties from {\tt NLO} cross sections in the {\tt SM} backgrounds are 
estimated to be 15\%. The effect of the choice of {\tt PDF} on signal efficiencies is evaluated 
using the {\tt CTEQ6.1M} {\tt PDF} error set~\cite{ref:cteq6} resulting in a 8\% uncertainty.
The uncertainty from {\tt MC} statistics can reach 10\% for the {\tt SM} and 5\% for signal.
The total uncertainty from isolated electron, muon, and track vetoes is 9\%. 
The uncertainty from heavy-flavor tagging is 12\% for {\tt SM} and  8\% for signal. 
Finally, the uncertainty from the trigger efficiency is 5\%.

\begin{figure}[hbtp]
\begin{minipage}{\linewidth}
  \begin{center}
   \includegraphics[width=1.\linewidth]
{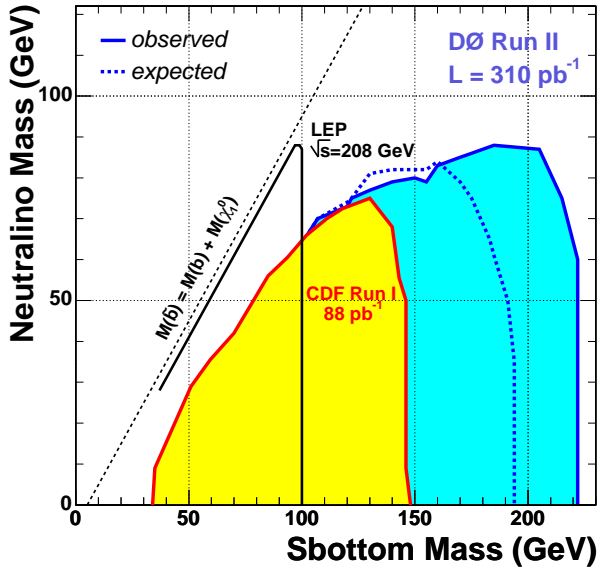}
  \end{center}
   \caption{Excluded regions at the 95\%~C.L. in the sbottom and neutralino mass plane.
The new region excluded by this analysis is shown in dark shading. The dashed line corresponds 
to the expected limit. Regions excluded by previous experiments are also displayed in the 
figure~\cite{theo:D0-CDF-LEP-sb}.} 
    \label{fig:Contour_Limitb}
\end{minipage}
\end{figure}

Since we do not observe any excess in the data relative to the expectations
from {\tt SM} backgrounds, we set limits on the production of sbottom quarks.
Observed and expected 95\% confidence level (C.L.) cross section upper limits are
obtained using the modified frequentist approach~\cite{ref:cls},
with correlations included between systematic uncertainties.
The {\tt NLO} $\tilde{b}$ pair production cross section is subject to theoretical
uncertainties arising from the {\tt PDF} and from the renormalization and
factorization scale choices. 
For a $\tilde{b}$ mass of 200 GeV, a 16\% {\tt PDF} uncertainty is evaluated 
using the {\tt CTEQ6.1M PDF} error set, and a 12\% uncertainty is found
by varying the scale by a factor of two up or down. 
For a given neutralino mass, a
sbottom mass limit is obtained where the cross section upper limit intersects the
production cross section reduced by these uncertainties combined in quadrature.
The results are summarized in the 95\%~C.L. exclusion contours displayed in
Fig.~\ref{fig:Contour_Limitb}. At higher sbottom masses, no events are observed
where about three are expected, leading to an observed limit more constraining than
expected.  

In summary, this analysis represents the first Tevatron Run II search for pair production of scalar bottom quarks.
The exclusion contour we obtain is substantially more restrictive than the ones published with Run I Tevatron 
data. With the current analysis using 310~\invpb, the maximum \MSB\ excluded is 222~\GeV, 
an improvement of more than 70~\GeV\ with respect to previous results, and the most 
restrictive limit on the sbottom mass to date.\\

%
We thank the staffs at Fermilab and collaborating institutions, 
and acknowledge support from the 
DOE and NSF (USA);
CEA and CNRS/IN2P3 (France);
FASI, Rosatom and RFBR (Russia);
CAPES, CNPq, FAPERJ, FAPESP and FUNDUNESP (Brazil);
DAE and DST (India);
Colciencias (Colombia);
CONACyT (Mexico);
KRF and KOSEF (Korea);
CONICET and UBACyT (Argentina);
FOM (The Netherlands);
PPARC (United Kingdom);
MSMT (Czech Republic);
CRC Program, CFI, NSERC and WestGrid Project (Canada);
BMBF and DFG (Germany);
SFI (Ireland);
The Swedish Research Council (Sweden);
Research Corporation;
Alexander von Humboldt Foundation;
and the Marie Curie Program.
%

\end{document}